\documentclass[12pt]{article}
\usepackage{hyperref}
\usepackage{amsmath}
\usepackage{graphicx}
\usepackage{amssymb} 
\newcommand{\Mat}[1]{{{\boldsymbol{#1}}}}

\def\be{\begin{equation}}
\def\ee{\end{equation}}
\def\bea{\begin{eqnarray}}
\def\eea{\end{eqnarray}}
\def\dd{\mathrm{d}}
\title{Dirac-type equations in a gravitational field, with vector wave function}

\author{
Mayeul Arminjon\\
\small\it CNRS (Section of Theoretical Physics).\\
\small\it At Laboratory ``Soils, Solids, Structures---Risks,"\\ \small\it CNRS \& University of Grenoble, Grenoble, France.
\footnote {\,Part of this work was done while the author was at Dipartimento di Fisica, Universit\`a di Bari and INFN Bari, Italy.} 
}
\date{ }

\begin{document}

\maketitle

\begin{abstract}
An analysis of the classical-quantum correspondence shows that it needs to identify a preferred class of coordinate systems, which defines a torsionless connection. One such class is that of the locally-geodesic systems, corresponding to the Levi-Civita connection. Another class, thus another connection, emerges if a preferred reference frame is available. From the classical Hamiltonian that rules geodesic motion, the correspondence yields two distinct Klein-Gordon equations and two distinct Dirac-type equations in a general metric, depending on the connection used. Each of these two equations is generally-covariant, transforms the wave function as a four-vector, and differs from the Fock-Weyl gravitational Dirac equation (DFW equation). One obeys the equivalence principle in an often-accepted sense, whereas the DFW equation obeys that principle only in an extended sense. 
\\


\end{abstract}

\section {Introduction}
Quantum mechanics in a classical gravitational field is of interest, for matter behaves undoubtedly as if it would obey quantum mechanics, while the gravitational field is undoubtedly there, and does behave classically as far as we know. Quantum effects for matter particles in the classical gravitational field are indeed being observed \cite{COW1975,WernerStaudenmannColella1979,RiehleBorde1991,KasevichChu1991,Nesvizhevsky2002}. The present work stems from the idea that the already-existing theoretical tools to analyse such effects are not necessarily the last word. Until recently, the analysis of relevant experiments has been based on the non-relativistic Schr\"odinger equation in the Newtonian gravity potential \cite{Nesvizhevsky2002,OverhauserColella1974,LuschikovFrank1978,VoroninAbele2006}, which is certainly not the last word: it is justified by the weakness of the gravitational field and by the smallness of the velocities involved. However, one expects that the precision of such measurements will increase significantly. Since neutrons are spin half particles, one may consider that, in the absence of a magnetic field, their behaviour should be correctly described by the Dirac equation. 
\footnote{\ 
For a neutral particle obeying Dirac's equation, the presence of an electromagnetic field has no effect whatsoever. This is contrary to the observed existence of a magnetic moment for the neutron, detected in magnetic scattering experiments, and interpreted as showing the neutron to be non-elementary.
}
In the case with a gravitational field, described by a curved spacetime, the Dirac equation is usually modified to the form derived independently by Weyl and by Fock in 1929, hereafter the {\it Dirac-Fock-Weyl (DFW)} equation. Thus, work has been done to study the physical consequences of the DFW equation, naturally in the framework of its weak-field and/or non-relativistic limit: see {\it e.g.} Refs. \cite{deOliveiraTiomno1962, Mashhoon1988, HehlNi1990, VarjuRyder1998, Obukhov2001}---but the corrections to the non-relativistic Schr\"odinger equation in the Newtonian gravity potential are usually very small. For instance, in the experiments on gravitational stationary states, one uses ultra-cold neutrons in the Earth's gravitational field \cite{Nesvizhevsky2002}. It has been shown recently that, in this particular case, the corrections brought by the DFW equation to the non-relativistic Schr\"odinger equation in the gravity potential are quite hopelessly negligible \cite{A38,Boulanger-Spindel2006}. Nevertheless, one may expect that, in the future, experiments (possibly using lighter neutral particles: massive neutrinos?) should be able to check this kind of corrections and, therefore, to distinguish between possible competing gravitational extensions of relativistic quantum mechanics.\\

The present work is a contribution to preparing theory to analyze such future experiments: we shall attempt to reconsider the formulation of quantum mechanics in a classical gravitational field and will derive alternative wave equations in this context. Prior to this, let us warn about a special feature of the two new gravitational Dirac-type equations to be derived here.

\subsection{Vector vs. spinor representation of the Dirac wave function}\label{TRD}

One particularity of the Dirac-type equations to be discussed here, is that both are based on the tensor representation of the Dirac fields (TRD): the wave function $\psi $ is a vector field on the spacetime, and the set of the four Dirac matrices $\gamma ^\mu$ builds a third-order $(^2 _1)$ tensor field, with components $\gamma ^{\mu \rho  }_\nu  \equiv (\gamma ^\mu )^\rho  _{\ \,\nu  }$ (writing, as usual, the row index $\rho $ of the Dirac matrices as a superscript and the column index $\nu $ as a subscript). That is, we get after a general change of coordinates, $x'^\mu =f^\mu ((x^\sigma  ))$  \cite{A37}:
\be \label{TRD-transform}
\psi'^\mu = \frac{\partial x'^\mu }{\partial x^\sigma  }\psi ^\sigma ,\qquad 
\gamma'^{\mu \rho  }_\nu= \frac{\partial x'^\mu }{\partial x^\sigma  }\frac{\partial x'^\rho  }{\partial x^\tau  }\frac{\partial x^\chi  }{\partial x'^\nu }\gamma ^{\sigma \tau }_\chi ,
\ee
or in matrix form:
\be \label{TRD-transform-matrix}
\Psi '=L\Psi, \quad \gamma '^\mu = L^\mu_{\ \, \sigma }\, L \gamma ^\sigma  L^{-1},
\ee
where
\be \nonumber
\Psi \equiv (\psi ^\mu ), \qquad L\equiv (L^\mu_{\ \,\sigma }),\quad L^\mu_{\ \,\sigma } \equiv \frac{\partial x'^\mu }{\partial x^\sigma  }.
\ee
This transformation leaves the {\it usual} flat-spacetime Dirac equation covariant after any linear transformation \cite{A37}, thus in particular after any Lorentz transformation, and even more particularly after any spatial rotation. It should be clear that the transformation (\ref{TRD-transform-matrix}) is as admissible as is the usual, spinor, transformation (which leaves the $\gamma ^\mu$ matrices invariant), from the viewpoint of special relativity. Indeed, an archetypical example of a relativistic transformation is that of the equation of motion for a charged test particle in an electromagnetic field $F^\mu _\nu $. In this equation, we have a coefficient matrix $F\equiv (F^\mu _{\ \,\nu} )$, in a very similar way as we have the four coefficient matrices $\gamma ^\mu $ in the Dirac equation---and the matrix $F$ is transformed thus:
\be \label{F'}
F'=L F L^{-1},
\ee
hence it is {\it not} invariant in a Lorentz transformation \cite{A37}. When a matrix enters a basic physical equation, it usually indicates the presence of some structure in the physical space or in the physical spacetime. The matrix is then the coordinate expression of some {\it tensor}. After, say, a spatial rotation of the coordinate system, the structure is seen rotated, and accordingly the matrix is not left invariant, but instead undergoes a transformation (\ref{F'}) or a similar one, depending on the tensor type. Besides that of the field tensor $F^\mu _\nu $ above, typical examples---by no means the only ones---are those of the stress and strain tensors in a deformed material. That the Dirac matrices stay invariant instead, when the spinor transformation is used, enforces one to consider that the wave function is a field, not in the physical spacetime, but in some ``internal space". This is mathematically expressed by the fact that a spinor belongs to a rather elaborated vector bundle, which leads to the relative mathematical complexity of the rigorous Dirac theory. In contrast, a spacetime vector is a familiar object in physics, {\it e.g.} the four-velocity. It belongs to just the tangent bundle to the spacetime, and its behaviour under a rotation has no mystery. Therefore, replacing spinors by vectors seems to go in the sense of physical understandability, mathematical simplicity, and an easier compatibility of Dirac theory with the rest of physics.\\

The obvious physical question is: does it threaten the experimentally-verified predictions of the Dirac equation? As recalled above, the gravitational Dirac equation is still essentially at a prospective stage (the existing experiments being well described by the non-relativistic Schr\"odinger equation), hence we can limit this question to the {\it flat-spacetime case}. The Dirac equation of special relativity is regarded as the correct relativistic quantum-mechanical wave equation for an elementary spin-half particle. It leads naturally to the emergence of spin-half states, {\it e.g.} its non-relativistic limit is the Pauli equation (see {\it e.g.} Ref. \cite{Schulten1999}). It allows the accurate calculation of the energy levels of hydrogen-type atoms ({\it e.g.} \cite{Schulten1999}). It also leads to the QED predictions associated with the corresponding Feynman propagator ({\it e.g.} \cite{BjorkenDrell1964}). As we observed \cite{A37}, these facts are derived from the validity of the Dirac equation in a given Cartesian coordinate system [usually with a particular set of gamma matrices $(\gamma ^\mu )$]. Thus, these derivations do not depend on whether the wave function is transformed (if one {\it changes} the coordinate system) as a spinor or as a vector. The energy eigenvalues and eigenfunctions, in particular, depend of course only on the explicit form of the equation in the given Cartesian coordinate system. Now this explicit form of the Dirac equation (also {\it with an electromagnetic field,} characterized by the potential $A_\mu $), as it is written matricially in any given Cartesian coordinates in a flat spacetime: 
\be \label{Dirac-em}
i\gamma ^\mu (\partial_\mu+iqA_\mu )\Psi  -m\Psi =0,
\ee
is left {\it unchanged} by our tensor representation. The two differences with the usual spinor representation are that, now, the column matrix $\Psi $ changes according to the simple rule (\ref{TRD-transform-matrix})$_1$  after a Lorentz transform, while the matrices $\gamma ^\mu$ do change according to Eq. (\ref{TRD-transform-matrix})$_2$. 
\footnote{\
Recall that the flat-spacetime Dirac equation (\ref{Dirac-em}) with spinor representation is limited to Cartesian coordinates, in contrast with the tensor representation (\ref{TRD-transform-matrix}). 
}
The latter point means that, to compare the predictions of the spinor ({\bf i}) and vector ({\bf ii}) representations in a given inertial reference system, we cannot {\it a priori} assume that the Dirac matrices are the same for ({\bf i}) and ({\bf ii})---since this equality could happen only by chance, being not maintained after a Lorentz transform. The question thus is whether there is any influence of the chosen set of Dirac matrices.\\

In precise terms: in a given Cartesian system, is there any influence of the chosen set $(\gamma ^\mu )$ of matrices solutions of the standard anticommutation relation
\be \label{Clifford-4}
\gamma ^\mu \gamma ^\nu + \gamma ^\nu \gamma ^\mu = 2\eta ^{\mu \nu}\,{\bf 1}_4, \qquad \mu ,\nu \in \{0,...,3\}
\ee
on the quantum-mechanical predictions obtained with the Dirac equation (\ref{Dirac-em})? To this question, the answer is a clear {\it no.} Indeed, a detailed study of the current 4-vector and its conservation, the Hilbert space inner product $(\Psi \parallel  \Phi )$, the hermiticity of the Hamiltonian H, and the transition amplitudes $(\mathrm{H}\Psi \parallel \Phi )$, has been done with a fully general set of matrices $(\gamma ^\mu )$ solution of (\ref{Clifford-4}) (and in fact in general affine coordinates: not necessarily Cartesian ones). It has thus been proved \cite{A40} that all of these elements are exactly {\it invariant} under a general change of the set: $(\gamma ^\mu )\mapsto (\tilde{\gamma ^\mu} )$ with $\tilde{\gamma ^\mu} =S\gamma ^\mu S^{-1}$ where $S$ is any matrix in $ {\sf GL(4,C)}$, accompanied by the corresponding change $\Psi \mapsto \tilde{\Psi }\equiv S\Psi $ for the wave function. \{Results making plausible the independence of Dirac theory on the set $(\gamma ^\mu )$, in the case of a unitary change: $S \in{\sf U(4)}$, had been obtained by Pal \cite{Pal2007}.\} We conclude that {\it there is simply no difference between the quantum-mechanical predictions of the Dirac equation (\ref{Dirac-em}) with vector or spinor representation of the Dirac wave function, in the domain of validity of the latter, {\it i.e.,} in Cartesian coordinates in a flat spacetime.} Inside this domain, differences might possibly occur only from a quantum-field-theoretical treatment, although the fully negative result found for quantum mechanics leads us to expect no such occurrence. The present work deals with quantum mechanics, but in a curved spacetime. We shall derive two wave equations that differ from the standard (DFW) version of the gravitational Dirac equation. We do expect that they will give different physical predictions, both from each other and as compared with the DFW equation.

\subsection{Classical-quantum correspondence vs. equivalence principle}

The standard wave equations in a curved spacetime are got by a ``covariantization procedure," inspired by the {\it equivalence principle.} Due to its non-local character, QM is, in a sense, incompatible with the equivalence principle whose formulation is purely local \cite{AldrovandiPereiraVu2005}. Thus, as is well-known, the predictions of QM depend on the mass $m$ of the particle. Nevertheless, there is a genuine sense in which the equivalence principle can be used as a tool to formulate QM in a gravitational field: this consists in trying to rewrite the flat-spacetime wave equation in a generally-covariant form, in such a way that the flat-spacetime equation is {\it automatically} recovered in a {\it (holonomic) local freely-falling frame, i.e.,} in a coordinate system in which, at the event $X$ considered, the metric tensor reduces to the standard form $\eta_{\mu \nu } $, with matrix $(\eta_{\mu \nu }) \equiv \mathrm{diag}(1,-1,-1,-1)$, and the {\it metric} connection vanishes. This is in fact the standard procedure to implement the equivalence principle in the formulation of the laws of nongravitational physics in a curved spacetime \cite{Will1993}. The physical justification of this procedure is that, in a holonomic local freely-falling frame, the equations of motion of a test particle, and of a continuous medium as well, coincide at $X$ with those in a flat spacetime: ``the gravitational field is locally suppressed." Hence, it makes sense to impose that, in a such frame and at this event $X$, other equations of physics---like, for example, the Maxwell equations or the Klein-Gordon equation---also coincide with their flat-spacetime version. As one applies that method to the Klein-Gordon (KG) equation, he obtains the so-called ``minimally-coupled" version of the generally-covariant KG equation (see {\it e.g.} Ref. \cite{VillalbaGreiner2001}): this is the case $\xi =0$ of the generally-covariant KG equation containing the arbitrary parameter $\xi$ which multiplies the scalar curvature \cite{BirrellDavies}. In this {\it genuine sense,} the equivalence principle does apply to get the gravitational KG equation. The same is true for the minimally-coupled Maxwell equations.\\

Writing explicitly the DFW equation involves the choice not only of a coordinate system (chart), but also of a tetrad field \cite{Weyl1929,BrillWheeler1957+Corr}. Moreover, the covariant derivatives in the DFW equation are defined using the ``spin connection," which is a non-metric one. In an {\it extended sense,} the equivalence principle may be said to apply if the flat-spacetime equation is recovered in coordinates (and, possibly, with an appropriate choice of the tetrad field) for which, at the event considered, the metric tensor reduces to the flat form $\eta_{\mu \nu }$ and the {\it relevant} connection vanishes. (The relevant connection is that which defines the covariant derivatives in the equation.) This applies to the DFW equation: it does coincide with the flat-spacetime Dirac equation in a coordinate system (combined with an appropriate choice of the tetrad) such that, at the event considered, the metric tensor reduces to the flat form $\eta_{\mu \nu }$ and the {\it spin} connection vanishes. 
\footnote{\
This is proved in Appendix \ref{DFW-not-EP}. It is known that any connection on the {\it tangent bundle} TV to a differentiable manifold V can be reduced to zero at a point by an appropriate choice of the coordinate system and the vielbein field \cite{Hartley1995,Iliev1996} (see also Refs. \cite{vdHeyde1975,GronwaldHehl1996}). But this result does not apply to the spin connection, which is a connection on the bispinor bundle, that is a vector bundle of a general kind, and different from TV. {\it E.g.} in Ref. \cite{Hartley1995}, the proof depends crucially on the fact that the bundle is TV, since it introduces autoparallels---a notion whose definition needs to consider a connection on TV.
}
However, it is only in a holonomic local freely-falling frame in the sense defined in the foregoing paragraph that the gravitational field is locally suppressed. In a such frame, the DFW equation {\it can} be reduced to the flat Dirac equation by a special choice of the tetrad field, but does not {\it necessarily} reduce to it (Appendix \ref{DFW-not-EP}). Therefore, in our opinion, the DFW equation does not obey the equivalence principle in the standard sense. Furthermore, in more general spacetimes, including torsion and non-metricity, the spin connection is not unique, even though a unique Dirac operator may be distinguished under certain requirements \cite{ReiflerMorris2005}. {\it In summary,} the application of the covariantization procedure to the Dirac equation with spinor transformation leads to introduce the spin connection and to write the DFW equation, which obeys the equivalence principle only in an extended sense.  \\

Therefore, one may consider that the existing approach to the formulation of QM in a gravitational field, and in particular to the gravitational Dirac equation, is not fully compelling. It is the less so as anyway the experimental results are not constraining today, since they are well described by the non-relativistic Schr\"odinger equation. An alternative, more fundamental approach to this problem, is to come back to wave mechanics, {\it i.e., to associate a wave operator with a classical Hamiltonian}---as is done to get the wave equations in a flat spacetime. In this way, an alternative form of the Dirac equation in a gravitational field was derived recently \cite{A37}. This derivation applies also to the KG equation (in fact, the method was first applied to the KG equation \cite{B15}). However, the gravitational Dirac-type equation derived in Ref. \cite{A37} was limited to the static case, and it does not obey the equivalence principle in the genuine sense defined above---although it has been noted that a {\it new} generally-covariant Dirac-type equation, {\it obeying the equivalence principle} in that genuine sense, can be written as a by-product of the method utilized \{Eq. (71) in Ref. \cite{A37}\}. \\

The aim of the present paper is to study in detail this wave-mechanical approach to the quantum wave equations in a gravitational field. We begin by summarizing our previous analysis of wave mechanics, adding important precisions as to its coordinate dependence and the way to resolve it. Then, using this approach, we derive two versions of the gravitational KG equation and two versions of the gravitational Dirac-type equation, and this in the general case. Finally, for each of these two gravitational Dirac equations, we derive the balance equation obeyed by the most natural 4-current. 

\section{Classical-quantum correspondence for the Klein-Gordon and Dirac equations in a gravitational field}

Instead of using the equivalence principle so as to adapt to gravity a wave equation originally derived for flat spacetime, one might think of applying directly the {\it classical-quantum correspondence.} By the latter, we mean here the process of associating a wave operator with a classical Hamiltonian: this process is also known as ``wave mechanics," and it appears to be a crucial step in the transition from classical to quantum mechanics. (Of course, the relation between classical and quantum theories involves other aspects, including statistical ones \cite{Khrennikov2005}.) As we shall recall below (Subsect. \ref{class-quant-KG}), there is indeed a classical Hamiltonian for the motion of a test particle in a gravitational field. However, canonical quantization will depend on coordinates, and in a curved spacetime it seems that there are no preferred coordinates---whereas Cartesian coordinates are preferred for a flat spacetime. The points that we made \cite{A37,A22} are that
		\begin{itemize}
\item the classical-quantum correspondence results from the purely mathematical correspondence between a wave operator and its dispersion equation, and from the wave-mechanical principle according to which the classical trajectories represent the skeleton of a wave pattern;
\item this analysis makes it clear that using the classical-quantum correspondence needs to identify a preferred equivalence class of coordinate systems; at least in a static gravitational field, a such class does exist.
\end{itemize}
To make the paper self-contained, we briefly explain these points in subsects. \ref{classical-quantum} and \ref{corres-prefers} below. We add important new remarks, in particular we now note that the data of the equivalence class is equivalent to that of an affine connection on spacetime. Moreover, we observe that one may actually identify {\it two} distinct classes of coordinate systems, thus two connections. Then we apply this analysis to the Klein-Gordon (Subsect. \ref{class-quant-KG}) and Dirac equations (Subsect. \ref{class-quant-Dirac}). \\

To avoid misunderstandings, let us explain in advance why {\it our use of special classes of coordinate systems is justified:} simply, one is allowed to use particular coordinate systems---as is done at many places in every physics textbook, and as is also done in the majority of the physics papers---provided the final result does not depend on the coordinate system, inside some general-enough class of systems. Thus, in special relativity, the Cartesian coordinate systems (in which the metric tensor has the standard form $\eta_{\mu \nu }$ at any point) make a such class. In a curved spacetime, of course, there are no such global Cartesian systems. Now, each of our two gravitational Klein-Gordon and Dirac-type equations is covariant under a {\it general} spacetime coordinate change, because each of them is written with the help of a connection on spacetime. It is true that one of the two connections is directly defined only if the coordinate system is adapted to the assumed preferred reference frame, Eq. (\ref{Delta-connection-etherframe}) below. The occurrence of a privileged reference frame is a physical possibility currently explored by several researchers, precisely in the context of gravitation: {\it cf.} {\it e.g.} the ``Einstein-aether theory" of gravitation, endowed with a dynamical preferred reference frame \cite{JacobsonMattingly2001}; {\it cf.} also this author's scalar theory \cite{A34,A35}. This physical possibility may turn out to be experimentally relevant to quantum mechanics in a gravitational field, or not. But it does make sense to be investigated, even though it does not fit ideally in the spacetime formalism (in the sense that it does not automatically lead to generally-covariant equations). Preferred-frame equations like Eq. (\ref{KG-ether}) can be rewritten in a generally-covariant form, as is done here by using the extension of the connection (\ref{Delta-connection-etherframe}) to a general coordinate system, but this necessarily involves referring to the preferred frame in some way.

\subsection{Analysis of the classical-quantum correspondence} \label{classical-quantum}

The first part of the classical-quantum correspondence is purely mathematical. Consider a linear partial differential equation of the second order, as is sufficient for quantum mechanics. In an arbitrary coordinate system or chart $\chi : X\mapsto (x^\rho ),\mathrm{U}\rightarrow {\sf R}^{N+1}$ (with $X$ the position in the extended configuration space V, of dimension $N+1$, and $\mathrm{U}$ an open subset of V), it has a local expression:
\be\label{wave-eqn}
P\psi \equiv \left[a_0((x^\rho )) +a_1^\mu ((x^\rho ))\partial _\mu +a_2^{\mu \nu } ((x^\rho ))\partial _\mu \partial _\nu \right] \psi=0,
\ee
where $\psi: \chi (\mathrm{U}) \rightarrow  {\sf C}^{m}$ is the local expression of the unknown \cite{Dieudonne1974}. The latter may be a complex scalar field ($m=1$) or have a vector, tensor or spinor nature as well. The manifold V may occur as a product: $\mathrm{V}=\mathsf{R}\times$M with  M an $N-$dimensional configuration space, but this is not necessary. From Subsect. \ref{class-quant-KG} below, V will be our 4-dimensional spacetime ($N=3$), and we will often adopt the corresponding language already now, but, before Subsect. \ref{class-quant-KG}, $N$ might be any positive integer. Conflating $\psi $ with $\psi \circ \chi ^{-1}$, and $a_0 $ with $a_0 \circ \chi ^{-1}$, etc., which is harmless for local questions, we may consider that $\psi $, $a_0$, etc., are defined on (an open subset of) the spacetime V itself. Let us look for ``local plane-wave" solutions \cite{A22}, {\it i.e.,} functions $\ \psi (X) = A\,\exp[ i\theta (X)]$, with $\theta (X)$ {\it real}, such that, at the event $X$ considered, $\partial_\nu K_\mu (X)=0$, where 
\be
 K_\mu \equiv \partial _\mu \theta \qquad (\mu = 0,..., N)
\ee
are the components of the wave covector ${\bf K}$: $\omega\equiv -K_0$ is the frequency, and the ``spatial" covector ${\bf k}$ with components $K_j \ (j=1,...,N)$ is the ``spatial" wave covector. [Obviously,   ${\bf k}$ transforms as a covector under a change of the ``spatial" coordinates $x^j\ (j=1,...,N)$.] Substituting a such $\psi $ into (\ref{wave-eqn}) leads to the {\it dispersion equation,} a polynomial equation for ${\bf K}$:
\be\label{dispersion-eqn} 
\Pi_X({\bf K})\equiv a_0(X) +i\,a_1^\mu (X)K_\mu +i^2a_2^{\mu \nu } (X)K_\mu K_\nu  =0. 
\ee
Clearly, the linear operator $P$ (\ref{wave-eqn}) and its dispersion function $(X,{\bf K})\mapsto \Pi_X({\bf K})$ 
(\ref{dispersion-eqn}) are in {\it one-to-one correspondence} \cite{A37,Whitham}. The existence of {\it real} solution covectors ${\bf K}$ to (\ref{dispersion-eqn}) is the criterion that decides whether (\ref{wave-eqn}) can be termed a {\it wave equation}. The inverse correspondence  [from (\ref{dispersion-eqn}) to (\ref{wave-eqn})] is
\be \label{Pi-to-P}
K_\mu  \rightarrow \partial _\mu/i, \qquad (\mu =0,...,N). 
\ee
Note that the definition of the dispersion function $\Pi_X({\bf K})$ (\ref{dispersion-eqn}) and its one-to-one correspondence with the operator $P$ (\ref{Pi-to-P}) are valid also for a {\it non-scalar} wave function, $\psi (X)\in {\sf C}^m$ with $m \ne 1$, the coefficients common to $P$ and $\Pi_X$ being then matrices with $m$ rows and $m$ columns \cite{A37}: $a_0=(a_{0 \ \beta }^\alpha)$, etc., so that $\Pi_X({\bf K})=(\Pi_X({\bf K})^\alpha _{\ \beta })$ is then an $m \times m$ matrix, too.\\

The {\it dispersion relation(s)}: $\ \omega =W({\bf k};X)$, fix the wave mode \cite{Whitham}. Each of them is a particular root of $\, \Pi_X({\bf K}) =0$, considered as an equation for $\ \omega\equiv -K_0$. (This makes sense only for scalar equations.) Witham's theory of dispersive waves \cite{Whitham} still contains the crucial observation that the propagation of $\ {\bf k}$ turns out to be ruled by a {\it Hamiltonian system}:
\be \label{Hamilton-W-k} 
\frac{\dd k_j}{\dd t}= -\frac{\partial  W}{\partial x^j},
\ee
\be \label{Hamilton-W-x} 
\frac{\dd x^j}{\dd t}= \frac{\partial  W}{\partial k_j} \qquad (t\equiv x^0, \ j=1,...,N).\\
\ee
This leads to the other part of the classical-quantum correspondence. The idea of de Broglie-Schr\"odinger's {\it wave mechanics} is that a classical Hamiltonian $H$ describes the skeleton of a wave pattern. Then, Eqs. (\ref{Hamilton-W-k})-(\ref{Hamilton-W-x}) suggest that the wave equation should give a dispersion $W$ with the same Hamiltonian trajectories as $H$. The simplest way to do that is to assume that $H$ and $W$ are proportional, $H=\hbar W$... {\it This gives first $ E=\hbar \omega$, $ \ {\bf p}=\hbar {\bf k}$. Then, substituting $K_\mu  \rightarrow \partial _\mu/i$ (\ref{Pi-to-P}), it leads to the correspondence between a classical Hamiltonian and a ``quantum" wave operator.} See Refs. \cite{A37,A22} for details. 

\subsection{The classical-quantum correspondence needs specifying a connection} \label{corres-prefers}

Thus far, a fixed system of coordinates $(x^\mu)$ has been assumed given on the (configuration-)space-time V. Yet we must allow for coordinate changes, which change the coefficients of operator $P$ (\ref{wave-eqn}) in this way:
\be\label{changeP}
a'_{0}=a_0,\quad a_{1}'^{\rho}= a_1 ^{\mu} \frac{\partial x'^{\rho} }{\partial x^{\mu}}+a_2^{\mu \nu} \frac{\partial ^2 x'^\rho }{\partial x^\mu \partial x^\nu }, \quad a_2'^{\rho \theta } = a_2^{\mu \nu} \frac{\partial x'^\rho }{\partial x^\mu} \frac{\partial x'^\theta  }{\partial x^\nu}.
\ee
(This follows from the transformation of the derivatives. We assume here that $\psi $ is, or transforms as, a scalar.) It results from (\ref{changeP}) that, at a given event $X \in \mathrm{V}$, the dispersion polynomial $\Pi_X$ (\ref{dispersion-eqn}) remains invariant iff
\be\label{Pi_X-invariant}
a_2^{\mu \nu} \frac{\partial ^2 x'^\sigma  }{\partial x^\mu \partial x^\nu }=0, \qquad \sigma \in \{0,...,N\}.
\ee
Also, if the ``local plane-wave" condition is satisfied at $X$, {\it i.e.,} $\partial_\nu K_\mu (X)=0$, then it is still so in the new coordinates iff
\be\label{K_mu,nu=0-invariant}
\frac{\partial ^2 x^\rho   }{\partial x'^\mu \partial x'^\nu }K_\rho =0, \qquad \mu ,\nu  \in \{0,...,N\}.
\ee
Hence, the only way to ensure that the dispersion polynomial (\ref{dispersion-eqn}) and the condition $\partial_\nu K_\mu (X)=0$ stay invariant, is to identify a particular class  $\mathcal{C}_X$ of coordinate systems connected by changes satisfying
\be \label{infinit-linear}
\frac{\partial ^2x'^\rho}{\partial x^\mu\partial x^\nu}=0 \qquad \mu ,\nu ,\rho  \in \{0,...,N\}
\ee
at the event $X((x_0^\mu))=X((x_0'^\rho))$ considered, and to admit only those coordinate systems that belong to this class \cite{A22}.
\footnote{\
One may also consider the case where the wave function $\psi $ is not a scalar but a {\it vector} [a $(^1 _0)$ tensor] with components $\psi ^\beta $. Then, at fixed $\alpha $ and for coordinate changes belonging to a class  $\mathcal{C}_X$, the matrix coefficients $a_{0 \ \beta }^\alpha, a_{1 \ \beta }^{\mu \alpha}, a_{2 \quad \beta }^{\mu \nu \alpha}$ [see after Eq. (\ref{Pi-to-P})] behave as tensors of the kind specified by the other indices: $\beta ,\mu ,\nu $, and thus the corresponding coefficients $\Pi_X({\bf K})^\alpha _{\ \beta }$ behave as covectors (again at fixed $\alpha $).
} 
It is easy to check that (\ref{infinit-linear}) defines an {\it equivalence relation} $\mathcal{R}_X$ between coordinate systems (charts $\chi: Y \mapsto (x^\mu)$) that are defined in a neighborhood of $X \in \mathrm{V}$:
\be\label{R_X}
\chi \ \mathcal{R}_X\ \chi ' \mathrm{\ iff\ }\frac{\partial ^2x'^\rho}{\partial x^\mu\partial x^\nu}=0 \mathrm{\ at\ }X((x_0^\mu))=X((x_0'^\rho)). 
\ee
Now we have the \\

\noindent {\bf Theorem 1.} {\it The data, at each point $X \in \mathrm{V}$, of an equivalence class $\mathcal{C}_X$ of charts modulo the relation (\ref{R_X}), is equivalent to the data of a unique torsionless connection on the differentiable manifold $\mathrm{V}$, such that, at any point $X \in \mathrm{V}$, its coefficients vanish in any chart $\chi \in \mathcal{C}_X$.} \\

{\it Proof.} ({\bf i}) Suppose we have a torsionless connection $\Delta $ on V. It is well-known that, for any point $X \in \mathrm{V}$, there is a chart $\chi $ such that the coefficients $\Delta ^\mu _{\rho \nu }$ of that connection vanish at $X$ ({\it e.g.} \cite{L&L}). If we change the chart to another one $\chi '$, the coefficients change by the rule:
\be\label{transform-connec-Greek}
\Delta '^\mu _{\nu \rho }=\frac{\partial x'^\mu}{\partial x^\sigma }\frac{\partial x^\theta}{\partial x'^\nu }\frac{\partial x^\tau}{\partial x'^\rho } \Delta ^\sigma _{\theta \tau }+ \frac{\partial x'^\mu}{\partial x^\sigma }\frac{\partial ^2x^\sigma }{\partial x'^\nu \partial x'^\rho }.
\ee
It follows immediately from (\ref{transform-connec-Greek}) that the new coefficients $\Delta '^\mu _{\nu \rho }$ remain all zero at $X$, iff $\chi $ and $\chi '$ belong to the same equivalence class, say $\mathcal{C}_X$, modulo $\mathcal{R}_X$. Thus, a torsionless connection does define, at any point $X$, an equivalence class of charts modulo the relation (\ref{R_X}), by the property that its coefficients vanish at $X$ in each member of $\mathcal{C}_X$. ({\bf ii}) Conversely, suppose that, for any $X \in \mathrm{V}$, we have an equivalence class $\mathcal{C}_X$. If it exists a connection as in the Theorem, Eq. (\ref{transform-connec-Greek}) shows that, in any possible chart $\chi '$, and at any point $X$, its coefficients must be given by
\be\label{transform-connec-from-C_X}
\Delta '^\mu _{\nu \rho }=\frac{\partial x'^\mu}{\partial x^\sigma }\frac{\partial ^2x^\sigma }{\partial x'^\nu \partial x'^\rho },
\ee
where $\chi : Y \mapsto (x^\rho )$ is any chart belonging to the class $\mathcal{C}_X$. Lemma 2 in Appendix \ref{Transitivity-Connec} proves that this defines indeed a unique connection. Equation (\ref{transform-connec-from-C_X}) and the symmetry of second derivatives show that it is torsionless. Q.E.D.\\

Thus, the mathematical correspondence $K_\mu  \rightarrow \partial _\mu/i$ (\ref{Pi-to-P}) needs to specify, at any point $X \in \mathrm{V}$, an equivalence class $\mathcal{C}_X$ of coordinate systems modulo (\ref{R_X}), and this is equivalent to specifying a torsionless connection on V: the class $\mathcal{C}_X$ is that of the systems in which the connection coefficients {\it vanish} at $X$. Therefore, the local plane-wave condition $\partial_\nu K_\mu (X)=0$, restricted to coordinate systems of the class $\mathcal{C}_X$, may be written equivalently in any coordinates as $D_\nu K_\mu (X)=0$ (which indeed seems {\it a priori} more correct: {\it cf.} the case of flat spacetime in curvilinear coordinates), and moreover, if $\psi $ is a {\it scalar,} the wave equation (\ref{wave-eqn}) and the correspondence (\ref{Pi-to-P}) may be rewritten with the partial derivatives $\partial_\mu $ replaced by the covariant derivatives $D_\mu $ w.r.t. the relevant connection.
\footnote{\
The foregoing sentence follows from remarks made by a referee of this paper, who suggested to start directly in this way. To link that way of doing with the one based on using partial derivatives, I was led to state Theorem 1. The remarks of the referee allowed also to make the covariant rewriting of the KG and Dirac equations transparent.
} 
However, if $\psi $ is not a scalar but a {\it vector,} then $D_\mu D _\nu \psi \ne \partial _\mu \partial _\nu \psi $ even at $X$ in a chart of the class $\mathcal{C}_X$. In that case, the dispersion polynomial (\ref{dispersion-eqn}) and the second-order wave operator (\ref{wave-eqn}) [with $D_\mu $ in the place of $\partial_\mu $] do {\it not} exchange by the substitution $K_\mu  \rightarrow D _\mu/i$---because $a_2^{\mu \nu } K _\mu K _\nu = a_2^{\mu \nu } K _\nu K _\mu $ but $a_2^{\mu \nu } D _\mu D _\nu \psi \ne a_2^{\mu \nu } D _\nu D _\mu \psi$. Nevertheless, if the operator (\ref{wave-eqn}) is for a vector $\psi $ but is first-order in fact (all $a_2^{\mu \nu } $ matrices being zero), then the correspondence (\ref{Pi-to-P}) may still be rewritten with $\partial_\mu $ replaced by $D_\mu $. Moreover, if one defines the local plane-wave condition as $D_\nu K_\mu (X)=0$ {\it a priori,} he gets the correspondence $K_\mu  \rightarrow D _\mu/i$, between the dispersion polynomial and the wave operator, {\it also if the connection has torsion}---but still only for a scalar second-order equation or a first-order vector equation. \\

In particular, if $\mathrm{V}$ is endowed with a pseudo-Riemannian metric $\Mat{g}$, a relevant connection is the Levi-Civita connection. The corresponding class $\mathcal{C}_X^1$ is that of the locally-geodesic coordinate systems (LGCS) at $X$ for $\Mat{g}$, {\it i.e.}, the ones in which the Christoffel coefficients of $\Mat{g}$ vanish at $X$:
\be\label{LGCS}
\Gamma^\mu _{\nu\,\rho}(X)=0, \qquad \mu,\nu, \rho \in \{0,...,N\}. 
\ee
But the classical-quantum correspondence also contains the correspondence between the classical Hamiltonian $H$ and the dispersion relation $W$, $H=\hbar W$. The definition of $W$ (by solving $\ \Pi_X({\bf K}) =0$ for $\ \omega\equiv -K_0$) isolates the chosen time coordinate $t \equiv x^0$ among other possible choices $x'^0=\phi ((x^\mu))$. In the same way, the data of a classical Hamiltonian $H({\bf p},{\bf x},t)$ does distinguish a special time coordinate $t$. (These two occurrences of a special time coordinate are bound together, of course, since the correspondence assumes that $H=\hbar W$.) {\it In general,} the wave equation thus associated with $H$ will not be covariant under a change of the time coordinate, except for pure scale changes $x'^0=ax^0$ \cite{B15}. Hence, {\it a priori}, what we should do would be to impose that only spatial coordinate changes are allowed, get a wave equation, and only then examine its behaviour under general coordinate changes \cite{B15}. To do that, we have to define in a natural way a class of coordinate systems exchanging by purely spatial and infinitesimally-linear changes:
\be\label{preferredchanges}
x'^0=ax^0,\quad x'^j=\phi^j ((x^k)),\quad \frac{\partial ^2x'^j}{\partial x^k\partial x^l}(X)=0.
\ee

\vspace{3mm}
\subsection{A preferred-frame connection}\label{preferred-frame}

In the case of a {\it static} Lorentzian
\footnote{\ 
By a ``Lorentzian metric on V" we mean a pseudo-Riemannian metric with signature $(1,N)$. The static character of $\Mat{g}$ is defined just below.
} 
metric $\Mat{g}$ on V, a class satisfying (\ref{preferredchanges}) does appear naturally: the class $\mathcal{C}_X^2$ of the coordinate systems adapted to the static character of the metric ({\it i.e.,} such that we have $g_{\mu \nu}=g_{\mu \nu}((x^j))$ and $g_{0j}=0$), and which are locally-geodesic, at the event $X$ considered [or, equivalently in that case, at its spatial projection ${\bf x}$, with components $x^j \ (j=1,...,N)$], for the {\it spatial} part $\Mat{h}$ of the metric:
\be\label{spatiallyLGCS}
G^j_{k\,l}(X)=0,\quad j,k,l \in \{1,...,N\},
\ee
with $G^j_{k\,l}$ the Christoffel symbols of the spatial metric. Indeed, the transformations between static-adapted coordinate systems are just those satisfying (\ref{preferredchanges})$_1$ and (\ref{preferredchanges})$_2$ \cite{A16},
\footnote{\ 
Thus, a static metric distinguishes a preferred reference frame.
}
while, in the same way as for Eq. (\ref{infinit-linear}) in the case of any pseudo-Riemannian metric $\Mat{g}$, the spatial coordinate systems that are LGCS for the spatial metric $\Mat{h}$ are exactly those which exchange by infinitesimally-linear spatial transformations, Eq. (\ref{preferredchanges})$_3$. \\

Now, let us consider a {\it general} Lorentzian metric $\Mat{g}$ on V, and assume that for some reason we dispose of a {\it preferred reference frame} E, including the data of a preferred time coordinate $T$ (up to a scale change). Thus, by definition, the coordinates that are {\it adapted to} E exchange by changes satisfying (\ref{preferredchanges})$_1$ and (\ref{preferredchanges})$_2$. Moreover, the spatial metric $\Mat{h}$ associated in the given frame E with the Lorentzian metric $\Mat{g}$ \cite{L&L} has then a privileged status, too. Hence, we may extend the definition of the class $\mathcal{C}_X^2$ to the case of a general Lorentzian metric $\Mat{g}$, by defining $\mathcal{C}_X^2$ as the class of the systems adapted to E and that are LGCS at $X$ for $\Mat{h}$, Eq. (\ref{spatiallyLGCS}). As for the class $\mathcal{C}_X^1$, two systems belonging to the class $\mathcal{C}_X^2$ must exchange by an infinitesimally-linear coordinate change (\ref{infinit-linear}). But, this time, the converse is not true: since the class $\mathcal{C}_X^2$ is restricted to purely spatial changes, not all changes satisfying (\ref{infinit-linear}) are internal to $\mathcal{C}_X^2$, in other words that class is smaller than an equivalence class of relation $\mathcal{R}_X$. Moreover, no LGCS {\it for $\Mat{g}$} is in general bound to the given frame E, because the local observer of the frame E is in general not in a ``free fall." Therefore, the classes $\mathcal{C}_X^1$ and $\mathcal{C}_X^2$ have in general no intersection. (Recall that two equivalence classes are either equal or intersection-free.) An exception is (for $N=3$, say) when the metric $\Mat{g}$ is flat and the frame E is an inertial frame: in that case, $\mathcal{C}_X^2$ is (strictly) contained in $\mathcal{C}_X^1$.\\

Theorem 1 above associates a unique connection $\Delta $ with the class $\mathcal{C}_X^2$ ---or, equivalently, with the unique equivalence class of relation (\ref{R_X}) that contains $\mathcal{C}_X^2$. That theorem characterizes the connection $\Delta $ as the one whose coefficients all vanish in any coordinate system which is adapted to E {\it and} which is an LGCS for $\Mat{h}$ at the relevant point $X$, Eq. (\ref{spatiallyLGCS}). We now prove a more useful characterization: $\Delta $ {\it is the unique connection which has the following expression in any coordinate system that is adapted to E} (but which is not necessarily an LGCS for $\Mat{h}$):
\be\label{Delta-connection-etherframe}
\Delta^{\mu} _{\rho \nu} \equiv \left\{ \begin{array}{ll} 0 &  \mathrm{if}\ \mu=0 \mathrm{\ or\ }\nu =0 \mathrm{\ or\ }\rho =0\\
& \\
 G^{j} _{l k} & \mathrm{if\ }\mu=j \ \mathrm{and\ }\nu=k\ \mathrm{and\ }\rho = l \in \{1,...,N\}.
\end{array} \right.
\ee
The coordinate systems adapted to E exchange by changes (\ref{preferredchanges})$_1$-(\ref{preferredchanges})$_2$. After any such change, the spatial connection $G ^j _{lk}$ transforms according to the usual rule, {\it i.e.,}
\be\label{transform-connec-Latin}
G '^j _{k l }=\frac{\partial x'^j}{\partial x^m }\frac{\partial x^n}{\partial x'^k }\frac{\partial x^p}{\partial x'^l } G ^m _{n p }+ \frac{\partial x'^j}{\partial x^m }\frac{\partial ^2x^m }{\partial x'^k \partial x'^l }.
\ee
It is easy to check that, as a consequence, a change (\ref{preferredchanges})$_1$-(\ref{preferredchanges})$_2$ gets the whole array $\Delta^{\mu} _{\rho \nu}$, as defined by (\ref{Delta-connection-etherframe}), change according to the rule of connections (\ref{transform-connec-Greek}). The charts (coordinate systems) adapted to the preferred frame E must be two-by-two compatible, and, taken together, they must cover the spacetime V: they make an ``atlas" $\mathcal{A}$. Thus, for an atlas of charts of the manifold V, the modification of the array $\Delta^{\mu} _{\rho \nu}$ due to the change of chart obeys the transformation law of a connection. It follows that we endow V with a unique connection in {\it defining} now the $\Delta '^{\mu}_{\nu \rho }$'s by (\ref{transform-connec-Greek}) for a {\it general} spacetime chart $\chi': X\mapsto (x'^{\mu})$. Indeed, Theorem 2 of Appendix \ref{Transitivity-Connec} states that: i) the $\Delta '^{\mu} _{\nu \rho }$'s, obtained thus, are independent of the chart $\chi: X \mapsto (x^\nu )$, belonging to the atlas $\mathcal{A}$, which is used in Eq. (\ref{transform-connec-Greek}); ii) moreover, the connection coefficients still transform according to the same rule (\ref{transform-connec-Greek}) from a general system to another one; iii) there is only one connection on V which obeys (\ref{transform-connec-Greek}) for any chart $\chi\in\mathcal{A}$ and for any general chart $\chi '$. Since the coefficients $\Delta^{\mu} _{\rho \nu}(X)$ of the connection just introduced all vanish in any coordinate system which is adapted to E and which is an LGCS for $\Mat{h}$ at the relevant point $X$, it follows that this is the connection $\Delta $, as we claimed.\\

\subsection{Klein-Gordon equation(s) in a gravitational field:\\derivation from the classical Hamiltonian} \label{class-quant-KG}

For a particle subjected to geodesic motion with a Lorentzian spacetime metric $\Mat{g}$, there is a classical Hamiltonian in the usual sense. To our knowledge, this result has been first got for the static case: in Ref. \cite{B15}, it has been shown that the classical energy of the particle \cite{L&L} is then a Hamiltonian, which may be expressed as a function of the canonical momentum ${\bf p}$ and the position ${\bf x}$ in the static reference frame as
\be \label{HamiltonStatic}
H({\bf p},{\bf x})= [g_{00}(h^{jk}p_j p_k c^2 +m^2c^4)]^{1/2}, \qquad (h^{jk}) \equiv (h_{jk})^{-1}
\ee
(the canonical momentum ${\bf p}$ being in fact the usual momentum \cite{B15}). Bertschinger \cite{Bertschinger1999} shows that there is a Hamiltonian in the general case. He recalls first that the following Hamiltonian defined on the 8-dimensional phase space:
\be\label{Hamilton-8dim}
\tilde{H}((p_\mu),(x^\nu))\equiv \frac{1}{2}g^{\mu \nu}((x^\rho)) p_\mu p_\nu \qquad (c=1),
\ee
has the geodesic lines as its trajectories. \{This result can be found in Misner {\it et al.} \cite{MTW}. Actually, it is a particular case of a theorem stating that free motion ($H=T$) in a Riemannian manifold is geodesic motion: Sect. 45 D in Ref. \cite{Arnold1976}.\} Bertschinger \cite{Bertschinger1999} notes that this Hamiltonian is inconvenient, because ``every test particle has its own affine parameter" (a similar remark is Note 8 in Ref. \cite{B15}). Then he defines a Hamiltonian in the usual sense ({\it i.e.,} depending on the position in the 6-dimensional phase space, and on the independent time $t$) by ``dimensional reduction," which is got simply by solving (\ref{Hamilton-8dim}) for $H=-p_0$ (see Sect. 45 B of Ref. \cite{Arnold1976}). Thus, that usual Hamiltonian $H$ depends on the coordinate system. However, Bertschinger \cite{Bertschinger1999} notes that $\tilde{H}=-\frac{1}{2}m^2$. Hence, in the most general case, $H\equiv -p_0$ satisfies the {\it generally-covariant} relation
\be \label{dispersion-geodesic}
g^{\mu \nu} p_\mu p_\nu -m^2 =0 \qquad (c=1)
\ee
[which is easily checked from Eq. (\ref{HamiltonStatic}) in the static case. We changed the sign of $\Mat{g}$ between Eqs. (\ref{Hamilton-8dim}) and (\ref{dispersion-geodesic}) to recover the signature $(+---)$.] \\

Therefore, we may apply the classical-quantum correspondence. First, assuming $\hbar=1$ for convenience, the wave-mechanical correspondence $H=\hbar W$, {\it i.e.} $ E=\hbar \omega$, $ {\bf p}=\hbar {\bf k}$, writes simply $p_\mu=K_\mu$, so that (\ref{dispersion-geodesic}) is actually the dispersion equation (\ref{dispersion-eqn}) of the wave equation which is searched for. Then, the mathematical correspondence (\ref{Pi-to-P}) gives immediately the scalar wave equation
\be\label{KG-brut}
(g^{\mu \nu} \partial_\mu \partial_\nu +m^2)\psi  =0 \qquad (\hbar=c=1).
\ee
However, as we found in Subsect. \ref{corres-prefers}, we must specify the class of coordinate systems in which we use the classical-quantum correspondence and in which we thus get Eq. (\ref{KG-brut}); then we may rewrite (\ref{KG-brut}) in general systems, using the corresponding connection provided by Theorem 1, as
\be\label{KG-general-connection}
(g^{\mu \nu} D_\mu D_\nu +m^2)\psi  =0 \qquad (\hbar=c=1).
\ee
In the case that we have a preferred frame E (which case includes the static case), we may assume that (\ref{KG-brut}) holds for the class $\mathcal{C}_X^2$, which corresponds to the connection $\Delta $, thus $D_\mu =\Delta _\mu $ in (\ref{KG-general-connection}). Using the expression (\ref{Delta-connection-etherframe}) of $\Delta $, we may then write (\ref{KG-general-connection}) explicitly, in coordinates adapted to E, as
\footnote{\ 
We assume moreover that the $g_{0j}$ components vanish in one system adapted to E (and hence in all of them), which means that E admits a global synchronization \cite{A16,L&L}.
} 
\be\label{KG-ether}
\Delta^{(\Mat{h})} \psi -\frac{1}{c^2 g_{00}}\frac{\partial^2\psi }{\partial T^2}-M^2\psi=0,
\ee  
where $T$ is the preferred time coordinate, $M\equiv \frac{mc}{\hbar}$, and where
\be\label{defs}
\Delta^{(\Mat{h})}\, \psi \equiv \psi^{\mid j }_{\quad \mid j } = \frac{1}{\sqrt{h}} \left(\sqrt{h}\, h^{jk}\psi_{,k}\right)_{,j},
\ee
adopting furthermore the notations
\be \label{spatial-covariant}
u^{j}_{\quad \mid k } \equiv  u^{j}_{\,, k}+ G^j _{lk} u ^l
\ee
and
\be\label{raised-covariant}
\psi^{\mid j } \equiv h^{jk} \psi _{\,, k}
\ee
for the covariant derivative with respect to the spatial metric $\Mat{h}$. 
\\

On the other hand, without any assumption on a preferred frame, we may also write (\ref{KG-brut}) for the class $\mathcal{C}_X^1$, made of the LGCS at $X$ for $\Mat{g}$: indeed, it happens that a (coordinate-dependent) Hamiltonian $H$ satisfying (\ref{dispersion-geodesic}) is available in any spacetime coordinate system, and all of these Hamiltonians describe the same motion (that along geodesics of the metric connection). This is a very particular situation for a classical Hamiltonian, and is the reason why our {\it a priori} strategy of specifying a preferred reference frame turns out to be not the only one possible. Then, $D_\mu $ is the Levi-Civita covariant derivative in Eq. (\ref{KG-general-connection}), and the latter may obviously be rewritten as the minimally-coupled generally-covariant KG equation:
\be\label{KG-minimcoupl}
\square^{(\Mat{g})} \psi + M^2 \psi =0, 
\ee
\be
\square^{(\Mat{g})} \psi \equiv \psi^{;\mu }_{\quad ;\mu } = \frac{1}{\sqrt{-g}} \left(\sqrt{-g}\, g^{\mu \nu }\psi_{,\nu }\right)_{,\mu },\ g \equiv \mathrm{det}(g_{\mu \nu }),\ (g^{\mu \nu }) \equiv (g_{\mu \nu })^{-1}.
\ee
Clearly, the two equations (\ref{KG-ether}) and (\ref{KG-minimcoupl}) are distinct (incompatible), unless metric $\Mat{g}$ is flat. This comes simply from the fact that they correspond to writing the classical-quantum correspondence in either of the two distinct classes of coordinate systems, $\mathcal{C}_X^2$ and $\mathcal{C}_X^1$ respectively.

\subsection{Dirac equation(s) in a gravitational field: derivation from the classical Hamiltonian} \label{class-quant-Dirac}

In contrast with what happens in the nonrelativistic case, the relativistic dispersion equation (\ref{dispersion-geodesic}) does not express the Hamiltonian $H=-p_0$ as an explicit polynomial in the canonical momentum ${\bf p}$ (with components $p_j, \ j=1,2,3$), or equivalently in the spatial wave covector ${\bf k}$, but instead as an implicit algebraic function of it \cite{A37}. This (second-order) algebraic relationship (\ref{dispersion-geodesic}) has the ``right" solution, say $H({\bf p};X)$, but it has also another solution, {\it e.g.} simply $-H({\bf p};X)$ if $g_{0j}=0$, and this other solution is inappropriate since it describes a different motion. Thus, (\ref{dispersion-geodesic}) has too much solutions. As a consequence, the associated wave equation, {\it i.e.} the KG equation [either (\ref{KG-ether}) or (\ref{KG-minimcoupl}), which coincide in the case of flat spacetime], also has too much solutions. Therefore, it is tempting to try a {\it factorization} of the dispersion equation associated with the algebraic relation (\ref{dispersion-geodesic}) \cite{A37}:
\be \label{Pi-factorization}
\Pi_X({\bf K})\equiv [g^{\mu \nu}(X) K_\mu K_\nu -m^2]{\bf 1}_\mathrm{A}= [\alpha(X) +i \gamma ^\mu(X) K_\mu][\beta(X) +i \zeta ^\nu(X)  K_\nu],
\ee
where ${\bf 1}_\mathrm{A}$ means the unity in some algebra A, which must be larger than the complex field $\mathsf{C}$ (and hence may be noncommutative), because a decomposition (\ref{Pi-factorization}) cannot occur in $\mathsf{C}$. Identifying coefficients in (\ref{Pi-factorization}), and applying the correspondence $K_\mu  \rightarrow \partial _\mu/i\ $ (\ref{Pi-to-P}) to ({\it e.g.}) the first of the two first-order polynomials on the r.h.s., leads \cite{A37} to the Dirac equation:
\be \label{Dirac-brut}
(i\gamma ^\mu \partial_\mu -m{\bf 1}_\mathrm{A})\psi =0,
\ee
where the objects $\gamma ^\mu (X) \in \mathrm{A}$ have to obey the anticommutation relation
\be \label{Clifford}
\gamma ^\mu \gamma ^\nu + \gamma ^\nu \gamma ^\mu = 2g^{\mu \nu}\,{\bf 1}_\mathrm{A}, \qquad \mu ,\nu \in \{0,...,N\},
\ee
with $N=3$ for our four-dimensional spacetime. This derivation works as it is summarized here, the metric $\Mat{g}$ being not necessarily the flat metric, and being instead a general Lorentzian metric on spacetime \cite{A37}. It works with a slight modification if one adds an electromagnetic potential $A_\mu$ \cite{A37}. As we know, it turns out that the algebra generated by any set $(\gamma ^\mu)$, solution of (\ref{Clifford}) with $N=3$, is isomorphic to that of $4\times4$ complex matrices, thus $\mathrm{A}= {\sf M}(4,{\sf C})$. Hence, $\psi (X)$ has four complex components $\psi ^\rho (X)\ (\rho =0,...,3)$. (See Footnote 9.)\\

As for the KG equation (\ref{KG-brut}) obtained from the classical-quantum correspondence, Eq. (\ref{Dirac-brut}) makes sense only in coordinate systems belonging to the identified class: either $\mathcal{C}_X^1$ or also, if a preferred reference frame E is available, the class $\mathcal{C}_X^2$---and it rewrites in a general system by replacing $\partial_\mu $ with the covariant derivative $D_\mu $ w.r.t. the connection associated by Theorem 1 with the class considered. If we consider the class $\mathcal{C}_X^1$, this is the Levi-Civita connection, so that (\ref{Dirac-brut}) rewrites in a general system as 
\be \label{Dirac-general-equivalence principle}
\left(i\gamma^\nu D_\nu -M\right)\psi =0,\qquad (D_\nu \psi )^\rho \equiv \psi ^\rho_{;\,\nu} \equiv \partial _{\nu}\psi^\rho + \Gamma ^\rho_{\sigma \nu }\psi^\sigma,
\ee
with $\Gamma ^\rho_{\sigma \nu }$'s the Christoffel symbols of $\Mat{g}$. Thus, we have now derived from wave mechanics Eq. (\ref{Dirac-general-equivalence principle}), which had been merely noted in passing in Ref. \cite{A37}. As it had been noticed there, {\it Eq. (\ref{Dirac-general-equivalence principle}) obeys the equivalence principle} in the genuine sense, since it coincides with the flat-spacetime Dirac equation in coordinates such that the {\it metric} connection vanishes at $X$ and such that $g_{\mu \nu }(X)=\eta_{\mu \nu }$.\\

Similarly, if we have a preferred frame E at our disposal, and if we apply the classical-quantum correspondence in the class $\mathcal{C}_X^2$ of coordinate systems, we may rewrite (\ref{Dirac-brut}) as
\be\label{Dirac-ether}
\left(i\gamma^{\nu} \Delta {_\nu} -M\right)\psi=0, 
\ee 
with
\be\label{Delta_nu-psi^mu}
(\Delta _{\nu} \psi)^{\rho} \equiv \psi^\rho _{,\,\nu}+\Delta^{\rho} _{\sigma  \nu} \psi^\sigma  = \left\{ \begin{array}{ll} \psi^{\rho}_{,\,\nu } &  \mathrm{if}\ \rho=0 \mathrm{\ or\ }\nu =0\\
& \\
 \psi^{j}_{\mid k } & \mathrm{if\ }\rho=j \ \mathrm{and\ }\nu=k \in \{1,2,3\},
\end{array} \right.
\ee
the equality (\ref{Delta_nu-psi^mu})$_2$ being a consequence of Eq. (\ref{Delta-connection-etherframe}) and being thus valid only, like the latter, in systems adapted to E. [Recall that $ \psi^{j}_{\mid k } $ is the covariant derivative with respect to the spatial metric $\Mat{h}$, Eq. (\ref{spatial-covariant}).] Using the explicit expression of the $\Gamma ^\rho_{\sigma \nu }$'s for a metric such that $g_{0 j}=0$ (see {\it e.g.} Ref. \cite{A16}, Subsect. 3.2), one checks that (\ref{Delta_nu-psi^mu}) coincides, in the static case, with the expression previously found, involving the $\psi ^\rho_{;\,\nu}$'s and other terms built with $\Mat{g}$, Eqs. (74)-(77) in Ref. \cite{A37}. This means that Eq. (\ref{Dirac-ether}) extends to the general case the gravitational Dirac-type equation previously derived from wave mechanics in the static case. Like the DFW equation, Eq. (\ref{Dirac-ether}) obeys the equivalence principle only in the {\it extended sense} defined in the Introduction, since it coincides pointwise with the flat-spacetime Dirac equation in coordinates (which always exist \cite{L&L}) such that the torsionless connection $\Delta $ vanishes at $X$ and such that $g_{\mu \nu }(X)=\eta_{\mu \nu }$---but, in general, this is {\it not} a local freely-falling frame: Eq. (\ref{LGCS}) is generally not true in these coordinates.\\

Thus, we have derived from wave mechanics two different versions of the Dirac equation in a gravitational field, Eqs. (\ref{Dirac-general-equivalence principle}) and (\ref{Dirac-ether}). However, this writing in a general coordinate system involves transforming $\psi $ as a {\it usual 4-vector} (though generally a complex one),
\footnote{\
At each event $X$, the vector $\psi (X)$, with components $\psi ^\rho (X)$,  is thus an element of the vector space $\mathrm{T}_{\sf C}\mathrm{V}_X$, that is the complexification of the tangent space $\mathrm{T}\mathrm{V}_X$ at $X \in \mathrm{V}$. This remark makes it more obvious that the corresponding index $\rho $ is not a ``spacetime index", in that it does not refer to the components of the point $X \in\mathrm{V}$, the {\it domain} of the field $\psi $, but to the components of the (generally complex) vector $\psi (X)$, that belongs to the {\it range} of this field. (See Footnote 4 in Ref. \cite{A42} for the ``index types" in other relevant tensors.) The possibility of this vector representation depends, as a referee noted, on the condition that the anticommutation relation (\ref{Clifford}) be satisfied by matrices $n\times n$ with $n=N+1$, the dimension of the spacetime V. As far as we know, $n=4$. Interestingly, $n=4$ might turn out to be the unique realistic possibility to satisfy this condition, for it seems from Ref. \cite{Pais1962} that matrices obeying the anticommutation relation (\ref{Clifford}) should generate the algebra ${\sf M(2^\nu ,C)}$ with $\nu$ the integer part of $\frac{n}{2}$. Now $n=2^\nu$ is verified only for $n=1,2$, and $4$. The first two are undoubtedly not the dimension of our spacetime! That is, the possibility of the vector representation would demand that the dimension of our spacetime be what it actually seems to be.
}
 because we are using the usual covariant derivative of a vector, Eqs. (\ref{Dirac-general-equivalence principle})$_2$ and (\ref{Delta_nu-psi^mu})$_1$. As recalled in Subsect. \ref{TRD}, the  flat-spacetime Dirac equation is Lorentz-covariant if one transforms the wave function as a 4-vector, provided that one simultaneously transforms the set of the $\gamma ^\mu$ matrices as a $(^2 _1)$ tensor, Eq. (\ref{TRD-transform}). With the transformation (\ref{TRD-transform}) applied to a general coordinate change, $(D_\nu \psi)^\rho$ as given by (\ref{Dirac-general-equivalence principle})$_2$, and $(\Delta _\nu \psi)^\rho$ in Eq. (\ref{Delta_nu-psi^mu}), are $\ (^1 _1)$ tensors, hence the l.h.s. of (\ref{Dirac-general-equivalence principle})$_1$, as well as that of (\ref{Dirac-ether}), is a {\it vector} under a general coordinate change. Hence, Eqs. (\ref{Dirac-general-equivalence principle}) and (\ref{Dirac-ether}) are generally-covariant.

\section{Balance equations for the new gravitational Dirac-type equations}

\subsection{Definition of the field of Dirac matrices} 
For the standard gravitational Dirac equation, which is the DFW equation, the field of the matrices $\gamma ^\mu (X)$ satisfying the anticommutation relation (\ref{Clifford}) is defined covariantly from an orthonormal tetrad $(u_\alpha )$, with $u_\alpha   \equiv  a^\mu  _{\ \,\alpha} \, \frac{\partial }{\partial x^\mu  } \quad (\alpha =0,...,3)$, by \cite{deOliveiraTiomno1962,VillalbaGreiner2001,BrillWheeler1957+Corr}
\be \label{flat-deformed}
\gamma ^\mu   = a^\mu_{\ \,\alpha}  \ \tilde{\gamma }^\alpha ,
\ee
where $(\tilde{\gamma }^\alpha )$ is a set of ``flat" Dirac matrices, that satisfies Eq. (\ref{Clifford}) with $(g^{\mu \nu })=(\eta _{\mu \nu })^{-1}=(\eta _{\mu \nu })\equiv \mathrm{diag}(1,-1,-1,-1)$. Under a change of coordinates, the tetrad $(u_\alpha )$ is (of course) left unchanged, hence the matrix $(a^\mu  _{\ \,\alpha})$ changes to
\be\label{change-a-matrix}
a'^\mu  _{\ \,\ \alpha} = \frac{\partial x'^\mu}{\partial x^\nu  }a^\nu  _{\ \,\alpha}.
\ee
We can apply this in the same way to our alternative equations (\ref{Dirac-general-equivalence principle}) and (\ref{Dirac-ether}). However, for the DFW equation, the flat matrices $\tilde{\gamma }^\alpha $ are left unchanged in a coordinate change. With (\ref{flat-deformed}) and (\ref{change-a-matrix}), this leads to the vector behaviour of the ``deformed" set $(\gamma ^\mu )$ \cite{BrillWheeler1957+Corr}. In contrast, for Eqs. (\ref{Dirac-general-equivalence principle}) and (\ref{Dirac-ether}), the array $\gamma^{\mu \nu} _\rho \equiv  \left(\gamma^\mu \right)^\nu _{\ \,\rho}$ is a $\ (^2 _1)$ tensor. This is got with (\ref{flat-deformed}) and (\ref{change-a-matrix}) by individually transforming each flat matrix $\tilde{\gamma }^\alpha $ as a $\left (^1 _1\right)$ tensor,
\be\label{change-flat-Dirac-matrix}
\left(\tilde{\gamma}'^\alpha \right)^\nu _{\ \,\rho}  = \frac{\partial x'^\nu}{\partial x^\tau}\frac{\partial x^\phi }{\partial x'^\rho } \left(\tilde{\gamma}^\alpha  \right)^\tau  _{\ \,\phi}.  
\ee
As needed, this preserves the anticommutation relation (\ref{Clifford}) for the $\tilde{\gamma }^\alpha $ matrices, with $\eta ^{\mu \nu }$ in the place of $g^{\mu \nu }$. In turn, this anticommutation relation implies, as usual, {\it i.e.,} independently of the transformation behaviour of the matrices $\tilde{\gamma }^\alpha $, the anticommutation (\ref{Clifford}) for the actual metric $g^{\mu \nu }$. \{This is checked by using (\ref{flat-deformed}) together with the orthonormality condition of the tetrad $(u_\alpha )$---or, more precisely: together with the orthonormality condition of the dual tetrad $(u^\alpha_* )$ [such that $u^\alpha_*(u_\beta)=\delta ^\alpha _\beta $], which follows from the latter.\}\\

Alternatively, one may also define the set $(\gamma ^\mu )$ of the deformed matrices simply by parallelly transporting the $\gamma^{\mu \nu} _\rho $ tensor along the geodesic lines of the metric connection associated with $g_{\mu \nu }$. Thus, let the $\gamma ^\mu$ matrices be defined in any way (perhaps from a local tetrad) at some event $X_0 \in \mathrm{V}$, of course satisfying (\ref{Clifford}) at this event $X_0$. Then, for another event $X$, let $G$ be the geodesic line (of the metric connection) that joins $X_0$ to $X$. We assume $G$ to be unique; this is always the case in some neighborhood U of $X_0$---the smaller the curvature, the larger U. With the aim to follow spin half particles in the real world, assuming this uniqueness seems to be safe. Transporting the $\gamma^{\mu \nu} _\rho $ tensor along $G$ does define matrices $\gamma ^\mu$ matrices satisfying (\ref{Clifford}): indeed, this may be rewritten as  
\be \label{Clifford-component}
\gamma^{\mu \rho } _\sigma  \gamma ^{\nu \sigma }_\tau  + \gamma^{\nu \rho } _\sigma  \gamma^{\mu \sigma  } _\tau  = 2g^{\mu \nu } \delta ^\rho _\tau, \qquad \mu ,\nu, \rho,\tau  \in \{0,...,3\}.
\ee
For any regular line, thus for $G$, there exists a coordinate system in which the connection coefficients $\Gamma ^\mu _{\nu \rho }$ vanish along this line \cite{Fermi1922,Cartan1951}. In a such coordinate system, parallelly transported tensors are simply constant along $G$, component by component: this applies, by construction, to $\gamma^{\mu \nu} _\rho $, and it applies to $g^{\mu \nu }$ since $g^{\mu \nu }_{;\sigma }=0$ implies that $g^{\mu \nu }$ is parallelly transported along any line. Hence, the equality (\ref{Clifford-component}), assumed to hold at $X_0$, holds true all along $G$ (in this system, hence in all, this being a tensor equation), hence everywhere in U.\\

Thus, we described two ways [namely, from a tetrad field or by parallel transport] to define a tensor field of Dirac matrices satisfying the anticommutation (\ref{Clifford-component}). This does not exhaust the possibilities.

\subsection{Balance equations for the usual current vector} 

The derivation of the current conservation for the flat Dirac equation ({\it e.g.} \cite{Schulten1999}) can be extended to the DFW equation \cite{Audretsch1974}. The corresponding definition of the current may be adapted to the alternative gravitational Dirac equations (\ref{Dirac-general-equivalence principle}) and (\ref{Dirac-ether}), but this current obeys only a balance equation with a source term---as we now show. Avoiding reference to a special set of the flat Dirac matrices, the standard derivation may be based on a ``hermitizing matrix" \cite{Pauli1933,Kofink1949}, {\it i.e.,} a Hermitian matrix $A =(A_{\rho \nu })$ such that each of the $\tilde{\gamma }^\alpha $ matrices is a Hermitian operator for the Hermitian  product $(u ,v ) \equiv A_{\rho \nu } u ^{\rho*} v^\nu$, which occurs iff we have 
\be\label{hermitizing-A}
A_{\rho \nu } \left(\tilde{\gamma}^{\alpha \, *} \right)^{\rho}  _{\ \,\sigma} = A_{\sigma  \rho }\left(\tilde{\gamma}^\alpha  \right)^{\rho  } _{\ \,\nu}  \qquad \alpha  ,\nu  ,\sigma  \in \{0,...,3\},
\ee 
Clearly, if the ``deformed" matrices $\gamma ^\mu$ are related to the $\tilde{\gamma }^\alpha $'s by a tetrad (as is always possible), Eq. (\ref{flat-deformed}), then $A$ is also hermitizing for the $\gamma ^\mu$'s:
\be\label{hermitizing-A-tensor}
A_{\rho \nu } \gamma^{\mu \rho *}  _\sigma = A_{\sigma \rho  }\gamma^{\mu \rho } _\nu  \qquad \mu, \nu  ,\sigma  \in \{0,...,3\}.
\ee
Note that this is a tensor equation. We define the $\gamma ^\mu$'s in that way at event $X_0$; for any other event $X$ we define the $\gamma^{\mu \nu}  _\rho (X)$ and $A_{\rho \nu }(X)$ tensors by parallel transport on the geodesic $G$, as above. We may define a 4-vector current by analogy with the flat Dirac and DFW equations:
\be\label{j-mu-standard}
j^\mu \equiv (\gamma ^\mu \psi ,\psi ) = A_{\rho \nu } \gamma^{\mu \rho *} _\sigma\psi ^{\sigma *} \psi ^{\nu },
\ee
and let us use again a coordinate system in which the Christoffel coefficients $\Gamma ^\mu _{\nu \rho }$ vanish along $G$. Then $A_{\rho \nu } =\mathrm{Constant}$ along $G$, hence (\ref{hermitizing-A-tensor}) still applies (in these, hence in any coordinates). In other words, the parallelly-transported hermitizing matrix defines a hermitizing tensor field. We have also $D_\mu=\partial_\mu$ along $G$, whence
\be\label{D_mu-jmu}
D_\mu j^\mu = A_{\rho \nu ;\mu } \gamma^{\mu \rho *}  _\sigma \psi ^{\sigma *} \psi ^\nu +((D_\mu \gamma ^\mu) \psi ,\psi )+(\gamma ^\mu D_\mu \psi ,\psi )+(\psi ,\gamma ^\mu D_\mu \psi ),
\ee
which again holds true in any coordinates, being a tensor equation. One may then use the relevant gravitational Dirac equation [either (\ref{Dirac-general-equivalence principle}) or (\ref{Dirac-ether})]: for instance, if this is Eq. (\ref{Dirac-general-equivalence principle}), the two last terms in (\ref{D_mu-jmu}) cancel one another. But anyway the r.h.s. remains with several source terms, which do not vanish individually nor cancel as a whole (being largely independent); {\it e.g.,} unlike $\partial_\mu \gamma ^\mu$ for the flat Dirac equation in Cartesian coordinates and $D_\mu \gamma ^\mu$ (with the covariant spinor derivative) for the DFW equation, we cannot expect that this $D_\mu \gamma ^\mu$ ({\it i.e.,} tensor $\gamma ^{\mu \nu }_{\rho ;\mu }$) vanish. [{\it Except for flat spacetime:} in that case, parallelly-transported tensors are constant in Cartesian coordinates, hence {\it the r.h.s. of (\ref{D_mu-jmu}) vanishes for a solution of either Eq. (\ref{Dirac-general-equivalence principle}) or (\ref{Dirac-ether}), which are equivalent and reduce to the flat-spacetime Dirac equation} (assuming the preferred frame is an inertial frame).] If there is a true conservation equation for either (\ref{Dirac-general-equivalence principle}) or (\ref{Dirac-ether}), and with the natural definition (\ref{j-mu-standard}) of the current, the tensor field $\gamma^{\mu \nu} _\rho $ should be defined in a different way than from parallel transport. This question is being investigated in detail in a forthcoming work \cite{A42}, where the current conservation is found to ask for definite field equations to be satisfied by the field $\gamma^{\mu \nu} _\rho $.

\section{Conclusion}

The effects combining relativistic gravity with relativistic quantum mechanics are hardly measurable for the time being, but they should become so in a not-too-distant future. This represents a window of observability for quantum gravity, and there are some reasons to believe that the basic equations used to analyze such effects might turn out not to be the most relevant ones. Thus, the standard (Dirac-Fock-Weyl or DFW) extension of the Dirac equation to gravitation is inspired by the equivalence principle, yet we find that it fails to obey this principle in the genuine sense described in the Introduction. (However, we agree that the DFW equation does obey the equivalence principle in an extended sense, see the Introduction.) \\

We start from an analysis of the classical-quantum correspondence, which makes that correspondence work also for a curved spacetime---whereas the way to the standard equations of relativistic quantum mechanics in curved spacetime has been to rewrite the flat-spacetime wave equations in generally-covariant form. In that way, the question how one should write (relativistic) QM in a (relativistic) classical gravitational field, becomes linked to the more general quest for an understanding of quantum mechanics. We find that the classical-quantum correspondence has to be written in special {\it classes} of coordinate systems, and that {\it two} distinct such classes may be identified. (The second class needs to distinguish a preferred reference frame.) These two classes lead to two distinct Klein-Gordon equations, Eqs. (\ref{KG-minimcoupl}) and (\ref{KG-ether}), and to two distinct Dirac-type equations in a curved spacetime, Eqs. (\ref{Dirac-general-equivalence principle}) and (\ref{Dirac-ether}). For flat spacetime, the second class is a subclass of the first one, and the equations coincide. Each of the two Dirac equations (\ref{Dirac-general-equivalence principle}) and (\ref{Dirac-ether}) is in fact generally-covariant.\\

In addition, for each of these two Dirac-type equations, {\it the wave function is a four-vector,} and the set of the Dirac matrices builds a $\left(^2 _1\right)$ tensor; see Subsect. \ref{TRD}. This tensor character means that the tetrad fields are at most an auxiliary tool in the new theory, whereas they play an essential role in DFW theory---as was first recognized by Weyl \cite{Weyl1929}. Note that, in our alternative equations, the tensor transformation behaviour (\ref{TRD-transform}) passes unchanged from Cartesian coordinates in a flat-spacetime to general coordinates in a general spacetime. In contrast, the transformation behaviour of the DFW equation under a coordinate change does not coincide with that for Dirac's original equation: for DFW, the bispinor is {\it invariant} under any coordinate change; spinor transformations act in DFW {\it only} as similarity transformations \cite{BrillWheeler1957+Corr}.\\

One of the alternative Dirac equations in a curved spacetime obeys the equivalence principle in the genuine sense, and the other one has a preferred reference frame---the latter might lead to larger corrections to the Schr\"odinger equation in the Newtonian potential. However, the current conservation, the Hamiltonian operator and its hermiticity, and their dependence on the possible tensor field of Dirac matrices, remain to be studied \cite{A42} for these two equations: in the present work, they have a tentative status. \\


\bigskip 
\noindent {\bf Acknowledgement.} I am grateful to Luc Rozoy for pointing out Ref. \cite{Cartan1951} and for interesting discussions on connections. Thanks are also due to Frank Reifler for his apt remarks on this work. The remarks of the three referees allowed to improve the paper significantly; this includes a technical improvement, see Footnote 5.
\bigskip 
\appendix
\section{The spin connection does not necessarily vanish in a local freely-falling frame}\label{DFW-not-EP}
The covariant spinor derivative is $D_\mu \equiv \partial _\mu -\Gamma _\mu$, with \cite{VillalbaGreiner2001,BrillWheeler1957+Corr}
\be \label{Gamma}
\Gamma _\mu = c_{\lambda \nu \mu }  s^{\lambda \nu}, \qquad c_{\lambda \nu \mu }\equiv \frac{1}{4} \left(g_{\lambda \rho }\, b^\beta _{\ \,\nu ,\mu} \,a^\rho  _{\ \,\beta}  - \Gamma _{\lambda \nu \mu}\right),
\ee
where $ s^{\lambda \nu} \equiv \frac{1}{2} \left (\gamma ^\lambda \gamma ^\nu - \gamma ^\nu \gamma ^\lambda \right )$, and the matrix $a=(a^\mu  _{\ \,\alpha} )$, with inverse $b=(b^\alpha  _{\ \,\mu})$, transforms the natural basis $(e_\mu)$ to the local tetrad $(u_\alpha )$, which is orthonormal, {\it i.e.,}
\be \label{ortho-tetrad}
\Mat{g}(u_\alpha  , u_\beta ) = a^\mu  _{\ \,\alpha}  \,a^\nu _{\ \,\beta}  \,g_{\mu \nu } = \eta _{\alpha \beta}, \qquad (\eta_{\alpha \beta}) \equiv \mathrm{diag}(1,-1,-1,-1) \equiv \mathrm{diag}(d_\mu),
\ee
which implies that
\be \label{g(e_mu,e_nu)}
g_{\mu \nu } = b^\alpha  _{\ \,\mu} \, b^\beta _{\ \,\nu} \,  \eta _{\alpha \beta  }.
\ee
The metric connection vanishes at event $X$ iff all derivatives $g_{\mu \nu ,\rho }$ vanish at $X$, or equivalently iff the first-kind Christoffel symbols $\Gamma _{\lambda \nu \mu}(X)$ are all zero. We assume from now that this condition is fulfilled and that, in addition, the metric tensor $g_{\mu \nu }(X)$ reduces to the flat form $\eta _{\mu \nu }$---thus, we have a ``(holonomic) local freely-falling frame"---and we check that, in general, (the other part of) $\Gamma _\mu(X)$ as given by Eq. (\ref{Gamma}) does not vanish then. Since $g_{\mu \nu }(X)=\eta _{\mu \nu }$, the orthonormality condition (\ref{ortho-tetrad}) is fulfilled by the natural basis $(e_\mu)$. Hence, the DFW equation being invariant under Lorentz transforms of the local tetrad \cite{BrillWheeler1957+Corr}, we may assume that
\footnote{\
Assuming (\ref{tetrad=natural}) just at the relevant event $X$, at which $g_{\mu \nu }(X)=\eta _{\mu \nu }$ in the chart utilized, we may restore the initial tetrad field, say $u'_\beta (Y)$, by a constant Lorentz transform $L$: $\forall Y \ u'_\beta (Y)=L^\alpha _{\ \,\beta}\, u_\alpha (Y) $, without changing the chart. This yields $a'=aL$, whence $b'_{,\mu }=L^{-1}b_{,\mu }$, thus leaving the $c_{\lambda \nu \mu }$'s unchanged in (\ref{Gamma}), whereas, owing to (\ref{tetrad=natural}), $s^{\lambda \nu}(X)$ becomes $s'^{\lambda \nu}(X)=L^\lambda _{\ \,\rho} L^\nu _{\ \,\sigma} s^{\rho \sigma }(X)$. Therefore, a matrix $\Gamma' _\mu = c_{\lambda \nu \mu }  s'^{\lambda \nu}$ vanishes at $X$ iff $\Gamma _\mu = c_{\lambda \nu \mu }  s^{\lambda \nu}$ already does.
}
\be\label{tetrad=natural}
a(X)=b(X)={\bf 1}_4,
\ee
which, by (\ref{g(e_mu,e_nu)}), ensures $g_{\mu \nu }(X)=\eta _{\mu \nu }$. With (\ref{tetrad=natural}), the cancellation of the $g_{\mu \nu ,\rho }$'s is equivalent, owing to (\ref{g(e_mu,e_nu)}), to
\be \label{b^nu_mu,rho}
b^\nu   _{\ \,\mu,\rho }\,d_\nu +  b^\mu  _{\ \,\nu ,\rho }\,d_\mu =0, \qquad \mathrm{no\ sum\ on\ }\mu \mathrm{\ and\ }\nu.
\ee
In turn, this is satisfied iff the following conditions hold for all $\rho \in \{0,...,3\}$:
\be \label{b^nu_mu,rho-detail1}
b^0_{\ \,0,\rho }=0, \qquad b^0_{\ \,j,\rho }=b^j_{\ \,0,\rho } \quad j \in \{1,2,3\},
\ee
and
\be \label{b^nu_mu,rho-detail2}
b^j_{\ \,k,\rho }=-b^k_{\ \,j,\rho }, \quad j,k \in \{1,2,3\}.
\ee
We get also from (\ref{tetrad=natural}), using (\ref{Gamma}) and the antisymmetry of $s^{\lambda \nu} $:
\be \label{Gamma-FFF}
\Gamma _\mu (X)=  \frac{1}{4} \sum_{\rho < \nu } d_\rho \left(b^\rho  _{\ \,\nu ,\mu}-b^\nu   _{\ \,\rho  ,\mu} \right) s^{\rho  \nu}. 
\ee 
Accounting for (\ref{b^nu_mu,rho}), we obtain easily from (\ref{Gamma-FFF}):
\be \label{Gamma-FFF-explicit}
\Gamma _\mu (X)=  -\frac{1}{2} \sum_{j < k} b^j  _{\ \,k ,\mu}\, s^{j k}. 
\ee
In addition to condition (\ref{b^nu_mu,rho}), that just rewrites $g_{\mu \nu ,\rho }=0$ when $a(X)={\bf 1}_4$, the only remaining constraint on the derivatives $b^\alpha _{\ \,\mu ,\rho }(X)$ is that the tetrad orthogonality (\ref{ortho-tetrad}) must hold at every event, and hence
\be \label{ortho-tetrad-deriv}
(a^\mu  _{\ \,\alpha}  a^\nu _{\ \,\beta}\,  g_{\mu \nu })_{,\rho } = 0.
\ee
However, our assumption [$a(X)={\bf 1}_4$ and hence $g_{\mu \nu }(X)=\eta _{\mu \nu }$] implies that (\ref{ortho-tetrad-deriv}) is equivalent (at $X$) to
\be \label{a^mu_alpha,rho}
a^\beta    _{\ \,\alpha ,\rho }\,d_\beta  +  a^\alpha   _{\ \,\beta  ,\rho }\,d_\alpha  =0, \qquad \mathrm{no\ sum\ on\ }\alpha  \mathrm{\ and\ }\beta.
\ee
The derivative of the inverse matrix $b=a^{-1}$ is $b_{,\rho }=-a^{-1}a_{,\rho }a^{-1}$, thus here $b_{,\rho }=-a_{,\rho }$, hence (\ref{a^mu_alpha,rho}) is a consequence of (\ref{b^nu_mu,rho}). In other words, Eq. (\ref{ortho-tetrad-deriv}) does not bring any additional constraint on the derivatives $b^\alpha _{\ \,\mu ,\rho }(X)$ in the present case. And since the constraint (\ref{b^nu_mu,rho}) is given explicitly by Eqs. (\ref{b^nu_mu,rho-detail1}) and (\ref{b^nu_mu,rho-detail2}), the derivatives $b^j  _{\ \,k ,\mu}(X) \quad (j<k)$ are left entirely free. It follows then from (\ref{Gamma-FFF-explicit}) that, indeed, {\it the spin connection matrices $\Gamma _\mu$ do not {\it usually} vanish ``in a holonomic local freely-falling frame," i.e.,} in a coordinate system such that $g_{\mu \nu ,\rho }(X)=0$ and $g_{\mu \nu }(X)=\eta _{\mu \nu }$ at the event $X$ considered. However, we {\it may always enforce} the spin connection to vanish in a such frame, by {\it choosing }  $b^j  _{\ \,k ,\mu}(X) =0 \quad (j<k)$ for all $\mu =0,...,3$, in addition to (\ref{b^nu_mu,rho-detail1}) and (\ref{b^nu_mu,rho-detail2})---that is, by imposing that all $b^\nu   _{\ \,\mu,\rho }$'s are zero at $X$. On the other hand, the $\gamma ^\mu$ matrices do coincide with the flat ones $\tilde{\gamma} ^\mu  $ in a local freely-falling frame. Hence, the DFW equation $i\gamma ^\mu(\partial _\mu -\Gamma _\mu)\psi =M\psi $ {\it can be made to coincide} with the flat Dirac equation $i\tilde{\gamma} ^\mu \partial _\mu \psi =M\psi $ in a such frame, but does not {\it usually} do so (except for special choices of the tetrad field).

\section{Definition of a connection on a manifold by extension from a smaller atlas of charts}\label{Transitivity-Connec}
What allows to define the connection $\Delta $ as explained after Eq. (\ref{transform-connec-Greek}) is essentially the transitivity property for the transformation rule of a connection:\\

\noindent {\bf Lemma 1.} {\it Let $\chi: X \mapsto (x^\mu ), \chi':X\mapsto (x'^\mu ),\tilde{\chi} :X\mapsto  (\tilde{x}^\mu )$ be three charts, two-by-two compatible, defined on the same open set $\mathrm{U}$ of a differentiable manifold $\mathrm{V}$, and let $\Delta^\mu _{\nu \rho },\Delta'^\mu _{\nu \rho },\tilde{\Delta}^\mu _{\nu \rho }$ be (point-dependent) arrays such that one goes from $\Delta $ to $\Delta'$ by Eq. (\ref{transform-connec-Greek}) and that the same transformation applies to go from $\Delta$ to $\tilde{\Delta}$, namely
\be\label{transform-connec-Greek-tilde}
\tilde{\Delta}^\sigma _{\theta \tau }=\frac{\partial \tilde{x}^\sigma}{\partial x^\phi }\frac{\partial x^\psi}{\partial \tilde{x}^\theta }\frac{\partial x^\zeta}{\partial \tilde{x}^\tau } \Delta ^\phi _{\psi \zeta }+ \frac{\partial \tilde{x}^\sigma}{\partial x^\phi }\frac{\partial ^2x^\phi }{\partial \tilde{x}^\theta \partial \tilde{x}^\tau }.
\ee
Then, the transformation rule of a connection still applies to go from $\tilde{\Delta}$ to $\Delta'$.}\\

 {\it Proof.} Define a new array $\tilde{\Delta}'^\mu _{\nu \rho }$ by just Eq. (\ref{transform-connec-Greek}), though substituting $\tilde{\Delta}$ for $\Delta$ and $\tilde{\Delta}'$ for $\Delta '$ (and substituting $\tilde{x}^\mu$ for $x^\mu$). We will show that in fact $\tilde{\Delta}'=\Delta '$, which shall prove the result. Inserting the $\tilde{\Delta}^\sigma _{\theta \tau }$'s given by (\ref{transform-connec-Greek-tilde}) into this definition of $\tilde{\Delta}'^\mu _{\nu \rho }$, we get (using implicitly the property that the three charts are two-by-two compatible at the $\mathcal{C}^2$ level and have the same domain of definition):
\be\label{Delta-tilde-prime}
\tilde{\Delta}'^\mu _{\nu \rho }=\frac{\partial x'^\mu}{\partial \tilde{x}^\sigma }\frac{\partial \tilde{x}^\theta}{\partial x'^\nu }\frac{\partial \tilde{x}^\tau}{\partial x'^\rho } \frac{\partial \tilde{x}^\sigma}{\partial x^\phi }\frac{\partial x^\psi}{\partial \tilde{x}^\theta }\frac{\partial x^\zeta}{\partial \tilde{x}^\tau } \Delta ^\phi _{\psi \zeta }+ \frac{\partial x'^\mu}{\partial \tilde{x}^\sigma }\frac{\partial \tilde{x}^\theta}{\partial x'^\nu }\frac{\partial \tilde{x}^\tau}{\partial x'^\rho }\frac{\partial \tilde{x}^\sigma}{\partial x^\phi }\frac{\partial ^2x^\phi }{\partial \tilde{x}^\theta \partial \tilde{x}^\tau }+ \frac{\partial x'^\mu}{\partial \tilde{x}^\sigma }\frac{\partial ^2\tilde{x}^\sigma }{\partial x'^\nu \partial x'^\rho }.
\ee
The second derivatives of composed functions are computed as
\be\label{compose-second-der}
\frac{\partial ^2x^\phi }{\partial x'^\nu \partial x'^\rho }=\frac{\partial \tilde{x}^\theta}{\partial x'^\nu }\frac{\partial \tilde{x}^\tau}{\partial x'^\rho }\frac{\partial ^2x^\phi }{\partial \tilde{x}^\theta \partial \tilde{x}^\tau } + \frac{\partial x^\phi }{\partial \tilde{x}^\sigma }\frac{\partial ^2\tilde{x}^\sigma }{\partial x'^\nu \partial x'^\rho }.
\ee
Inserting this into (\ref{Delta-tilde-prime}) yields
\bea \label{Delta'=Delta'tilde}
\tilde{\Delta}'^\mu _{\nu \rho }& = & \frac{\partial x'^\mu}{\partial x^\phi } \frac{\partial x^\psi}{\partial x'^\nu }\frac{\partial x^\zeta}{\partial x'^\rho } \Delta ^\phi _{\psi \zeta } + \frac{\partial x'^\mu}{\partial x^\phi }\left(\frac{\partial ^2x^\phi }{\partial x'^\nu \partial x'^\rho }-\frac{\partial x^\phi }{\partial \tilde{x}^\sigma }\frac{\partial ^2\tilde{x}^\sigma }{\partial x'^\nu \partial x'^\rho }\right) + \frac{\partial x'^\mu}{\partial \tilde{x}^\sigma }\frac{\partial ^2\tilde{x}^\sigma }{\partial x'^\nu \partial x'^\rho }\nonumber\\
& = & \frac{\partial x'^\mu}{\partial x^\phi } \frac{\partial x^\psi}{\partial x'^\nu }\frac{\partial x^\zeta}{\partial x'^\rho } \Delta ^\phi _{\psi \zeta } + \frac{\partial x'^\mu}{\partial x^\phi } \frac{\partial ^2x^\phi }{\partial x'^\nu \partial x'^\rho } = \Delta'^\mu _{\nu \rho },
\eea
which proves the Lemma.\\

 \noindent {\bf Theorem 2.} {\it Suppose that, on $\mathrm{V}$, there is an atlas $\mathcal{A}$ such that, for each chart $\chi \in \mathcal{A}$, a (point-dependent) array $\Delta ^\phi _{\psi \zeta }$ is defined, and that, for any two charts $\chi,\tilde{\chi} \in \mathcal{A}$, the arrays $\Delta ^\phi _{\psi \zeta }$ and $\tilde{\Delta}^\sigma _{\theta \tau }$ are related by Eq. (\ref{transform-connec-Greek-tilde}). Then, for any chart $\chi'$ belonging to the maximal atlas $\mathcal{M}$ compatible with $\mathcal{A}$, let us define $\Delta'^\mu _{\nu \rho }$ by Eq. (\ref{transform-connec-Greek}). This makes sense and defines a connection on $\mathrm{V}$ (for its manifold structure defined by $\mathcal{M}$),
\footnote{\ 
We adopt the elementary definition of a connection (on a differentiable manifold, more precisely on its tangent bundle) as an object given, in any chart, by a point-dependent threefold array which transforms under a change of chart according to the rule (\ref{transform-connec-Greek}), thus allowing to define a covariant differentiation of tensors \cite{Doubrovine1982}.
} 
which is the unique connection on $\mathrm{V}$ such that its coefficients $\Delta ^\phi _{\psi \zeta }$ are known for any chart $\chi \in \mathcal{A}$.}\\

 {\it Proof.} i) Let $\mathrm{U}$$'$ be the domain of $\chi '$ and let $X \in \mathrm{U}'$. Since $\mathcal{A}$ is an atlas, there is a chart $\chi \in \mathcal{A}$ such that its domain U contains $X$, and since $\mathcal{M}$ is the maximal atlas compatible with $\mathcal{A}$, the charts $\chi $ and $\chi '$ are compatible. Hence, we may use Eq. (\ref{transform-connec-Greek}) to define $\Delta'^\mu _{\nu \rho }$ on $\mathrm{U}\cap\mathrm{U}'$. If another chart $\tilde{\chi} \in \mathcal{A}$, with domain $\tilde{\mathrm{U}}$, is such that $\mathrm{U}\cap\mathrm{U}'\cap\tilde{\mathrm{U}} \ne \emptyset$, the proof of the Lemma shows that the definition of $\Delta'^\mu _{\nu \rho }$ from the chart $\tilde{\chi}$ (or rather from the transition map $\chi '\circ \tilde{\chi}^{-1}$) gives the same result as its definition from the chart $\chi $, Eq. (\ref{Delta'=Delta'tilde}). Thus $\Delta'^\mu _{\nu \rho }$ is well-defined on $\mathrm{U}'$.\\

ii) Let us now denote by $\tilde{\chi}$, not any more a chart in the small atlas $\mathcal{A}$, but instead another general chart (as $\chi'$): $\tilde{\chi} \in \mathcal{M}$, with domain $\tilde{\mathrm{U}}$. Thus the foregoing paragraph provides us unambiguously with the two arrays $\Delta'^\mu _{\nu \rho }$ and $\tilde{\Delta}^\sigma _{\theta \tau }$, {\it i.e.}, in the general coordinate systems $(x'^\mu)$ and $(\tilde{x}^\mu)$. Assume that $\mathrm{U}'\cap\tilde{\mathrm{U}} \ne \emptyset$, and let $X \in \mathrm{U}'\cap\tilde{\mathrm{U}}$. As above, there is a chart $\chi \in \mathcal{A}$ such that its domain U contains $X$, and the charts $\chi $, $\tilde{\chi}$ and $\chi '$ are compatible. Since $\chi \in \mathcal{A}$, we dispose of the array $\Delta^\mu _{\nu \rho }$ in the coordinate system $(x^\mu)$. As shown in the foregoing paragraph, one gets the array $\Delta'^\mu _{\nu \rho }$ from the array $\Delta^\mu _{\nu \rho }$ by Eq. (\ref{transform-connec-Greek}), {\it i.e.,} by the transformation rule of a connection, and the same rule applies to go from $\Delta^\mu _{\nu \rho }$ to $\tilde{\Delta}^\sigma _{\theta \tau }$. The Lemma shows that the transformation rule of a connection still applies to go from $\tilde{\Delta}$ to $\Delta'$ at any point $Y(x^\mu)=Y(x'^\mu)=Y(\tilde{x}^\mu) \in \mathrm{U}\cap\mathrm{U}'\cap\tilde{\mathrm{U}}$, hence in particular at $X$, thus at any point $X \in \mathrm{U}'\cap\tilde{\mathrm{U}}$. Thus, we have indeed defined a connection on V.\\

iii) The uniqueness of the connection results, obviously, from the fact that it {\it must} transform according to Eq. (\ref{transform-connec-Greek}), which is precisely used to define it. \\

The following Lemma is used in the proof of Theorem 1 in Subsect. \ref{corres-prefers}.\\

\noindent {\bf Lemma 2.} {\it Suppose that, at each point $X \in \mathrm{V}$, one has an equivalence class $\mathcal{C}_X$ of charts modulo the relation (\ref{R_X}). In any possible chart $\chi '$ (i.e.} $\chi '$ belongs to the maximal atlas  $\mathcal{M}$ that defines the manifold structure of $\mathrm{V}$), {\it and for any $X $ in the domain of $\chi '$, define coefficients by Eq. (\ref{transform-connec-from-C_X}), where $\chi : Y \mapsto (x^\rho )$ is any chart belonging to the class $\mathcal{C}_X$. The coefficients (\ref{transform-connec-from-C_X}) do not depend on the choice of the chart $\chi'  \in \mathcal{C}_X$ and transform by the rule (\ref{transform-connec-Greek}), thus they define a unique connection on $\mathrm{V}$. }\\

{\it Proof.} Let us first fix the point $X \in \mathrm{V}$ and the chart $\chi \in \mathcal{C}_X$. Consider another general chart  $\tilde{\chi} \in \mathcal{M}$, like $\chi '$, and assume that its domain also contains $X$. Define the coefficients, at $X$ and in the chart $\tilde{\chi}$, of the searched connection, just like Eq. (\ref{transform-connec-from-C_X}) does at $X$ and in the chart $\chi '$:
\be\label{transform-connec-from-C_X-tilde}
\tilde{\Delta}^\sigma _{\theta \tau }=\frac{\partial \tilde{x}^\sigma}{\partial x^\phi }\frac{\partial ^2x^\phi }{\partial \tilde{x}^\theta \partial \tilde{x}^\tau }.
\ee
Nearly the same calculation as Eqs. (\ref{Delta-tilde-prime})--(\ref{Delta'=Delta'tilde}), though in that simpler case, shows that we have
\be\label{transform-connec-tilde-to-prime}
\Delta '^\mu _{\nu \rho }=\frac{\partial x'^\mu}{\partial \tilde{x}^\sigma }\frac{\partial \tilde{x}^\theta}{\partial x'^\nu }\frac{\partial \tilde{x}^\tau}{\partial x'^\rho } \tilde{\Delta} ^\sigma _{\theta \tau }+ \frac{\partial x'^\mu}{\partial \tilde{x}^\sigma }\frac{\partial ^2\tilde{x}^\sigma }{\partial x'^\nu \partial x'^\rho }.
\ee
[Note, however, that here we are using coefficients $\Delta $ {\it only at point $X$}.] In particular, if in fact $\tilde{\chi} \in \mathcal{C}_X$, we have $\tilde{\Delta}^\sigma _{\theta \tau }=0$ by (\ref{transform-connec-from-C_X-tilde}), hence (\ref{transform-connec-from-C_X}) and (\ref{transform-connec-tilde-to-prime}) give:
\be\label{transform-connec-unique}
\Delta '^\mu _{\nu \rho }\equiv \frac{\partial x'^\mu}{\partial x^\sigma }\frac{\partial ^2x^\sigma }{\partial x'^\nu \partial x'^\rho }=\frac{\partial x'^\mu}{\partial \tilde{x}^\sigma }\frac{\partial ^2\tilde{x}^\sigma }{\partial x'^\nu \partial x'^\rho },
\ee
showing that indeed the coefficients (\ref{transform-connec-from-C_X}) do not depend on the choice of the chart $\chi  \in \mathcal{C}_X$. Thus, we have the coefficients $\Delta '^\mu _{\nu \rho }$ uniquely defined at any point $X \in \mathrm{V}$ and for any chart $\chi' \in \mathcal{M}$ whose domain contains $X$. Now, Eq. (\ref{transform-connec-tilde-to-prime}) proves that, on changing the chart $\chi' \in \mathcal{M}$, the coefficients change according to the rule of connections. Q.E.D.


\bigskip

\end{document}